\documentclass[conference]{IEEEtran}
\pdfoutput=1
\usepackage[hyphens]{url}
\usepackage{hyperref}
\hypersetup{colorlinks,allcolors=blue}
\usepackage[backend=biber, style=ieee]{biblatex}

\usepackage{amsmath,amssymb,amsfonts}
\usepackage{algorithmic}
\usepackage{graphicx}
\usepackage{textcomp}
\usepackage{colortbl}
\usepackage{xcolor}
\usepackage{adjustbox}

\usepackage{microtype}
\usepackage{array, booktabs}
\usepackage{graphics}
\usepackage{verbatim}
\usepackage{balance}
\def\BibTeX{{\rm B\kern-.05em{\sc i\kern-.025em b}\kern-.08em
    I\kern-.1667em\lower.7ex\hbox{E}\kern-.125emX}}

\begin{document}

\title{A Systematic Review and Taxonomy for Privacy Breach Classification: Trends, Gaps, and Future Directions}

\author{\IEEEauthorblockN{Clint Fuchs}
\IEEEauthorblockA{\textit{The Beacom College of Computer and Cyber Sciences} \\
\textit{Dakota State University }\\
Madison, USA \\
clint.fuchs@trojans.dsu.edu}

\and

\IEEEauthorblockN{John D. Hastings}
\IEEEauthorblockA{\textit{The Beacom College of Computer and Cyber Sciences} \\
\textit{Dakota State University }\\
Madison, USA \\
john.hastings@dsu.edu}
}

\maketitle

\begin{abstract}
In response to the rising frequency and complexity of data breaches and evolving global privacy regulations, this study presents a comprehensive examination of academic literature on the classification of privacy breaches and violations between 2010-2024. Through a systematic literature review, a corpus of screened studies was assembled and analyzed to identify primary research themes, emerging trends, and gaps in the field. A novel taxonomy is introduced to guide efforts by categorizing research efforts into seven domains: breach classification, report classification, breach detection, threat detection, breach prediction, risk analysis, and threat classification. An analysis reveals that breach classification and detection dominate the literature, while breach prediction and risk analysis have only recently emerged in the literature, suggesting opportunities for potential research impacts. Keyword and phrase frequency analysis reveal potentially underexplored areas, including location privacy, prediction models, and healthcare data breaches. 
\end{abstract}


\section{Introduction}

Cybersecurity breaches continue to be a global problem with 2023 seeing the highest volume of breaches of any of the previous 17 years and the number of attacks exploiting vulnerabilities as the main initiator increasing 180\% from the previous year \cite {Hylender2024}. \cite {Page_2024} supported those findings, reporting an overall increase in the volume of breaches as the threat landscape continues to evolve. Considering the massive amounts of data that are collected about each and every person \cite{allen2016protecting}, the findings should be a concern for every individual or entity that provides or collects sensitive data.

Privacy laws such as the European General Data Protection Regulation (GDPR), the California Privacy Rights Act (CPRA), and other emerging global regulations are reshaping how organizations must handle data. Hefty fines are incurred by data collectors who do not abide by the appropriate laws \cite{Wolford}. In 2023 alone, Meta was fined USD 1.3 billion (Facebook) and USD 413 million (Instagram) for GDPR infractions \cite{Hill_Sharma_2025}. This is just the tip of the iceberg. With the growing global adoption of privacy laws, organizations no longer have a choice; they are required to emphasize the protection of sensitive personal information that they collect, process, and store \cite{pedrosa2019gdpr}.

As regulatory requirements and the volume of collected data increase, so does the need for a comprehensive understanding of how privacy breaches occur and are classified in academic literature. In response, this study aims to develop a taxonomy to shed light on existing research involving the classification of privacy breaches and violations. The study is guided by the following research questions:

\begin{itemize}
\item RQ1: What are the trends in existing research related to the classification of
privacy breaches (violations)?
\item RQ2: Concerning privacy breach (violation) classification,
what are the under-researched areas and emerging
fields of study?
\item RQ3: At what frequency do select keywords (privacy,
breach, violation, etc.) occur within the surveyed literature?
\end{itemize}

The remainder of this paper is presented as follows. Section \ref{background} provides background related to privacy and privacy breaches. Section \ref{design} presents the methodology of the literature review. Section \ref{findings} presents the findings including the resulting novel taxonomy, followed by extended findings in Section \ref{extended}. Future work appears in Section \ref{future} with conclusions in Section \ref{conclusion}.

\section{Background}\label{background}

\subsection{Taxonomy}
A taxonomy is defined strictly as the classification of organisms, both living and extinct, or more succinctly, biological classification \cite {Cain_2025}. In a broader sense, taxonomy refers to the science of classification by establishing a hierarchy of superior and subordinate groupings \cite {Cain_2025}. Taxonomies give structure and organization to various fields of knowledge, allowing researchers to investigate relationships between concepts, thus allowing researchers to form hypothesis about these relationships \cite {fiedler1996empirically,glass1995contemporary,nickerson2013method}. In addition, taxonomies are helpful for the advancement of discussion, research, and pedagogy \cite{miller1994taxonomy,nickerson2013method}. \cite {Klaper_Hovy_2014} noted that the cybersecurity field is fragmented, with a broad scope. Many fields including cybersecurity benefit from the use of taxonomies to provide order where it may be lacking or nonexistent \cite {mcknight2002developing,nickerson2013method}.

\subsection{Privacy}
In the late 1800s, in response to newspaper circulation of photographs of private individuals, \cite {warren1890right} distilled privacy into ``the right to be let alone.'' Today, however, the boundaries of privacy have shifted dramatically. The rise of digital platforms has transformed privacy concerns from simple physical intrusions into issues of data surveillance and personal information management. As noted by \cite {solove2015meaning}, practically every business and organization with which we interact collects large amounts of personal data from all aspects of our lives, from the purchases we make, to the web sites we visit, and more. Individuals or data providers trade personal information as currency with service providers or data collectors in exchange for services \cite {auxier2019key}. To get an idea of the scope, Facebook \cite {boatwright2020privacy} collects over 52,000 unique attributes in each individual account holder. Personal information is the currency Facebook and other providers take as payment for the use of the services they provide.

\subsection{Privacy Breaches}
A general definition of a data breach is given by \cite{seh2020healthcare} as ``an illegal disclosure or use of information without authorization". The Department of Health and Human Services adds privacy and security to the definition of confidential health information \cite{wikina2014caused,seh2020healthcare}. \cite {bambauer2013privacy} posited that legal scholars merge privacy and security, while asserting that these are separate concerns. Data security and privacy are undoubtedly intertwined, and privacy is often viewed through the lens of security. Despite network security and privacy policies being in place, privacy is not guaranteed and can still be violated \cite {bhattacharya2006utilizing}. \cite{jouini2014classification} noted that studies tend to focus on the development of threat and harm taxonomies as they relate to security. These taxonomies are then applied to the classification of security breaches with a focus on mitigating the risk the classified breaches pose or the outcome specific security breaches incur.

\subsection{Privacy Legislation}
In 2018, California introduced the California Consumer Privacy Act of 2018 (CCPA) \cite {de_la_Torre_2018}. In the same year, the European Union (EU) adopted the General Data Protection Regulation (GDPR), which is considered by many to be the gold standard \cite{de_la_Torre_2018}. These laws serve to protect the citizenry of California and EU respectively. This protection goes beyond the territorial limits of each entity and has caused organizations all over the world to update data security and compliance due to the potentially large fines that can be incurred \cite {de_la_Torre_2018}.

GDPR Privacy Principles \cite {de_la_Torre_2018,Wolford}:
\begin{enumerate}
    \item Processing must be lawful, fair, and transparent to the EU citizens as data subjects.
    \item Data can only be processed for clearly stated legitimate reasons to the data subject.
    \item Data minimization must be in practice.
    \item Data accuracy and timeliness must be adhered to.
    \item Storage limitation and purpose limitation on information in that it can only be kept for the least possible time.
    \item All processing of data and data flows must ensure security, confidentiality, and integrity.
    \item The data controller is accountable for compliance with GDPR protection privacy laws.
\end{enumerate}

\section{Study Design}\label{design}

\subsection{Methodology}
The objective of this study is to discover existing and emerging trends and research gaps by conducting a systematic review of the literature on the classification of privacy breaches. To answer RQ1, RQ2, and RQ3, this study employed a quantitative approach in the implementation of a systematic literature review that was based on quantitative descriptive research methodology. This methodology was chosen due to it being non-experimental, as opposed to experimental designs, which focus on the measurement of selected variables. The non-experimental approach was a fit due to its ability to identify causality and that it is also possible to identify associations and relationships \cite{baker2017quantitative}.

\subsection{Literature Search}
To conduct the literature search, the terms of the primary research question (RQ1) were used to generate queries based on the terms and synonyms, as shown in Table \ref{tab:query_terms}. A corpus of literature was established through a systematic review of the literature utilizing several databases. These included the ACM Digital Library, IEEE Xplore, and Google Scholar. The ACM Digital Library and IEEE Xplore were selected on the basis of reputation and subject relevancy. Google Scholar was also included to increase the scope of the search and to include additional sources,including Elsevier and Springer. The search was conducted in November 2024.

\begin{table}[ht]
    \centering
       \caption{Literature Search Query Terms}
    \begin{tabular}{|c|c|}
    \hline
        security breach classification & cyber breach \\
        security breach classification & breaches categorization \\
        data security classification & privacy violations \\ 
        data privacy classification & threat classification \\ 
        privacy taxonomy breach & breach classification \\
        privacy framework violation &  privacy classification \\
        privacy framework data breach &privacy systematic \\
    \hline
    \end{tabular} \\
 
    \label{tab:query_terms}
\end{table}

\subsection{Literature Selection}
The set of articles found in the search was then narrowed using the inclusion and exclusion criteria listed in Table \ref{tab:ex-inc-criteria}. The screening process was conducted sequentially from the first result, until the studies resulting from the search were deemed unrelated according to the inclusion/exclusion criteria. This process was repeated for each search query term. The first round of the screening process eliminated candidates that met the exclusion criteria. The remaining candidates were manually selected by reviewing the abstract according to the inclusion criteria. The goal was to select papers that engaged in some form of classification related to data/privacy breaches or violations. This result was the exclusion of research whose focus did not match RQ1, such as studies involving data classification instead of breaches or violations.

\begin{table}[ht]
    \centering
    \caption{Study Inclusion/Exclusion Criteria}
     \begin{tabular}{|p{0.45\linewidth} | p{0.40\linewidth}|}
    \hline 
        \textbf{Inclusion Criteria} & \textbf{Exclusion Criteria} \\
    \hline 
        Studies involve classification/categorization and some combination of the terms data, privacy, breach, and/or violation & Study does not give a clear connection to RQ1 \\
        \hline
        In English & Not in English     \\
        \hline
        PDF downloadable & PDF unconvertible for analysis    \\ 
        \hline
        Relevant to topic (RQ1) & Secondary sources such as magazines or reviews     \\ 
        \hline
        Found in conference/journal & No duplicates         \\
    \hline
    \end{tabular} \\
    \label{tab:ex-inc-criteria}
\end{table}

\subsection{Taxonomy Construction}
After the selection process, a thematic analysis was performed on the final set of studies. From this analysis, a novel taxonomy was constructed to classify the focus areas of existing research related to privacy breach and violation classification. 

\subsection{Extended Corpus Analysis}
To complement the thematic taxonomy, additional corpus-wide analysis was conducted to uncover trends over time and frequency-based insights. This included keyword frequency counting, temporal distribution analysis of publication dates, and TF-IDF scoring across the selected literature set. These steps were taken to address RQ1–RQ3 in greater depth and are detailed in the extended findings section.

\section{Findings}\label{findings}

The literature search produced a large number of matches ranging from 400,000 to more than a million results for each query. From those, the targeted selection process produced a total of 29 articles published between 2010 and 2024: 23 in Google Scholar and six in IEEE Xplore with two of those articles appearing in (Table \ref{tab:database_selected_studies}). Google Scholar produced results from: ACM, AIS Electronic Library, Elsevier, HICSS, Little Lion Scientific, MDPI, Oxford University Press, River Publishers, Springer, and two that could not be determined. No papers were produced from the ACM Digital Library as the results focused primarily on classifying the privacy level of data or on aspects of differential privacy and thus did not meet the inclusion criteria.

\begin{table}[ht]
    \centering    
    \caption{Database and Selected Studies Count}
     \begin{tabular}{|p{0.3\linewidth} | p{0.32\linewidth}|}
    \hline 
        \textbf{Database} & \textbf{Selected Study Count} \\
    \hline 
        ACM Digital Library& 0 \\
    \hline
        Google Scholar & 23    \\
    \hline
        IEEE Xplore & 6 \\   
        
    \hline
    \end{tabular} \\

    \label{tab:database_selected_studies}
\end{table}

\subsection{Taxonomy}
The purpose of this study was to explore the classification of privacy breaches, but only a limited number of studies \cite {ayres2010standardizing,banerjee2011quantifying,rani2024data} from the selected corpus that met expectations for RQ1. During the analysis of the selected studies, the areas of focus found in the corpus revealed themselves, as shown in Table \ref{tab:taxonomy}. This resulted in a novel taxonomy for research related to privacy breach/violation classification that warrants future exploration. The related categories were further distilled into a macro set of privacy research focus categories for academic and potential future work in this field.

\begin{table}[ht]
    \centering
    \caption{Privacy Research Focus Taxonomy}
     \begin{tabular}{|p{0.28\linewidth} | p{0.55\linewidth}|}
    \hline 
        \textbf{Derived Categories} & \textbf{Distilled Categories} \\
    \hline 
        Breach Classification & Classification [Breaches, Reports, Threats] \\
          Detect Breaches & Detect [Breaches, Threats]    \\
          Detect Threats & Predict [Breaches, Threats, Outcomes] \\ 
          Predict Breaches & Risk [Breaches, Violations, Threats]     \\ 
          Report Classification &        \\

        Risk &        \\

        Threat Classification &        \\        
        
    \hline
    \end{tabular} \\

    \label{tab:taxonomy}
\end{table}

\subsubsection{Classification of Breaches} This category is important because security professionals learn mitigation strategies from documented breaches. To accomplish this task, researchers used machine learning to classify breaches according to a corresponding taxonomy. \cite {ayres2010standardizing,flak2019privacy,gore2018privacy,shah2021towards} focused on the classification of breach incidents. While \cite {gore2018privacy,seh2020healthcare} focused on the classification of IoT and IoMT data breaches, respectively. \cite {goh2023predictive} study focused on predicting the types of cyber-breach outcomes, a major portion of the study focused on developing a taxonomy and framework.

\subsubsection{Classification of Reports} 
This category involved machine learning for the classification of breach reports produced by organizations after breach events. The research conducted by \cite {blakely2022exploring,rani2024data} focused on the analysis of data breach reports and classification to improve internal controls. 

\subsubsection{Detect Breaches} 
This category consists of studies that leverage classification for the purpose of detecting breaches. \cite {ali2022big} defined four categories of privacy violations to focus on privacy challenges, privacy violations, and privacy preserving techniques. In relation to data stored on database systems, and \cite {silva2020using} leveraged NLP to personally identifiable information (PII) in data that was publicly available. \cite {banerjee2011quantifying} defined a four-dimensional taxonomy to record privacy violations, the severity of individual privacy violations, and the probability that a data provider would stop providing data due to violations and their severity. \cite {kokciyan2016priguardtool} developed a tool to detect privacy violations found on social networks. While \cite {luo2021automatic} developed a tool to detect privacy violations in Android applications. Studies focused on associated privacy violations, such as those associated with location-based services on mobile devices \cite {poikelarole} or via proximity-based applications \cite {puglisi2015potential}. \cite {umar2020study} developed detection models to identify self-disclosures of personal information.

\subsubsection{Detect Threats} 
Detecting potential threats to sensitive data or security gives security professionals direction for mitigation strategies. Studies in this category utilize classification for the purpose of identifying threats to sensitive data or security. \cite {islam2022smartvalidator} created a tool for the purpose of identification and classification to validate alerts and incidents. Similarly, \cite {raghunath2022xgboost} developed a tool to detect and classify cyber-attacks. \cite {pakhari2020implementation} instead focus on classification of unstructured data to conduct threat analysis. Insider threat detection was presented by \cite {meng2018deep}. \cite {wernke2014classification} focused on classification of location privacy attacks and approaches. 

\subsubsection{Predict Breach}
Predicting data breaches is an important undertaking that allows organizations to prepare for an event. If one is aware of potential security threats, defenses can be implemented. \cite {kumari2021prediction} used classification algorithms to predict network security threats by gathering pertinent attributes, including network protocol and service. \cite {li2024precursor} instead looked to predict privacy leakages from user-generated content.

\subsubsection{Risk}
Determining the risk associated with a particular activity or application provides security professionals with insight into which areas to focus resources.  \cite {uddin2023data} analyzed android apps by leveraging machine learning techniques to classify and identify Android applications that intentionally or unintentionally leak information to third parties.

\subsubsection{Threat Classification} 
This category is important because it quantifies threats, giving security professionals direction for the application of resources to combat. \cite {jouini2014classification} focus on threat class classification. The authors present a detailed multidimensional classification model that includes both internal and external threats.

\subsection{Data Extraction}
After the creation of the novel taxonomy depicted in the previous section, the corpus was manually reviewed by the author. This process involved identifying keywords from each study’s title, abstract, and methodology. Then each study was placed in a category based on fit. The first phase classified each study according to the first set of more fine-grained categories shown in Table \ref{tab:taxonomy} to facilitate RQ2. The next phase began by distilling the fine-grained categories into a macro set, also shown in Table \ref{tab:taxonomy} to get a broader overview of the corpus and trend identification for RQ1.

\subsection{Taxonomic Classification}

Table \ref{tab:selected_corpus} shows the classification of the articles in the corpus according to the taxonomy. \cite{liu2012automatic} note that taxonomies for specific domains are highly useful for a multitude of applications, and a taxonomy can be extracted from the key words and phrases found in a corpus. More importantly, \cite{liu2012automatic} stated that beyond simple keywords and phrases, additional knowledge and context is required to build a taxonomy. This serves as inspiration for the method used in this research. The selected studies that comprise the corpus, each provide key words and phrases that state the focus of the research. The context resulted in the categories that were derived from the corpus shown in Table \ref{tab:taxonomy}. The derived categories were distilled into macro categories that share high-level keywords and themes that match one of the distilled categories shown in Table \ref{tab:taxonomy}, allowing a high-level view of the research focus contained within the selected corpus. 

\begin{table}[h!]
   \caption{Study Corpus Classified by Macro Categories}
    \centering
      \begin{tabular}{|p{0.27\linewidth} | p{0.17\linewidth}| p{0.12\linewidth}| p{0.10\linewidth}|  p{0.07\linewidth}|}
    \hline 
        \textbf{Literature} & \textbf{Classification} & \textbf{Detect} & \textbf{Predict} & \textbf{Risk} \\
    \hline
        \cite {ali2022big,ayres2010standardizing,flak2019privacy,goh2023predictive,gore2018privacy,seh2020healthcare,shah2021towards,windl2023understanding,tran2023and} & Breach &  &  &  \\
        \hline
        \cite{ayoade2018automated,blakely2022exploring,rani2024data} & Report &  &  &  \\
        \hline
        \cite{jouini2014classification}  & Threat &  &  &   \\
        \hline
        \cite {banerjee2011quantifying,kokciyan2016priguardtool,luo2021automatic,panou2017riski,poikelarole,puglisi2015potential,silva2020using,umar2020study} &  & Breaches &  & \\
        \hline
        \cite{islam2022smartvalidator,meng2018deep,pakhari2020implementation,raghunath2022xgboost,wernke2014classification} &  & Threats &  &  \\
        \hline
        \cite{kumari2021prediction,li2024precursor} &  &  & Breach &   \\
        \hline
        \cite{uddin2023data} &  &  &  & Risk \\
    \hline
    \end{tabular} \\

    \label{tab:selected_corpus}
\end{table}

\section{Extended Findings and Analysis}\label{extended}

With the corpus in hand, a deeper analysis was conducted to comprehensively address the research questions and provide a broader understanding of current research. This extended analysis includes temporal trends, keyword frequency, and TF-IDF scoring across the selected literature.

\begin{figure}
    \centering
    \includegraphics[width=0.7\linewidth]{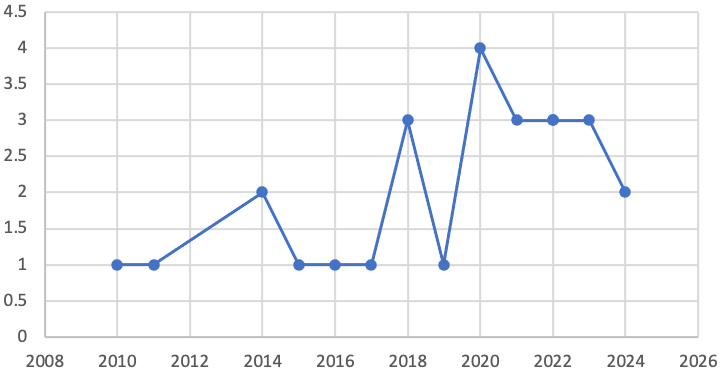}
    \caption{Corpus Published by Year}
    \label{fig: corpus-by-year}
\end{figure}

To understand how related research has progressed over time, Fig. \ref{fig: corpus-by-year} shows the number of studies found in the selected corpus by year. Initially, a 10-year window was selected for analysis, but with minimal results garnered from the past ten years, two outliers from 2010 and 2011 were included. It is interesting to note that the trend is in an upward trajectory. The data suggest that the increase in studies related to privacy classification corresponds to the adoption of GDPR in 2018 and COVID-19 in 2020, as noted by \cite {flak2019privacy,kumari2021prediction,silva2020using}. However, an investigation of causality is outside the scope of this study.

\subsection{RQ1: Classification Trends}
\begin{figure} [ht]
    \centering
    \includegraphics[width=0.9\linewidth]{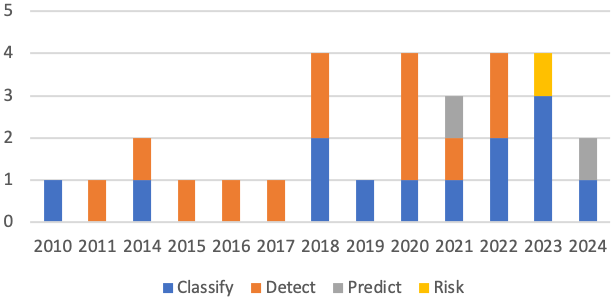}
    \caption{Publications Classified by Macro Categories by Year}
    \label{fig: macro-categories-by-year} 
\end{figure}

To address RQ1, the distilled taxonomy shown in Table \ref{tab:taxonomy} was used to classify the corpus. Fig. \ref{fig: macro-categories-by-year} reveals the results of classifying the corpus by year and macro categories. The data show that over the past 10+ years, the Classify and Detect categories of privacy or data breaches dominate the landscape. Fig. \ref{fig: macro-categories-by-year} shows that these two categories have been active throughout this period of time. However, Detect failed to make an appearance over the last two years, indicating that this area has likely been saturated. Classify, on the other hand, remains active. Predict and Risk do not appear until 2021 and 2023 respectively, indicating recent interest by researchers.

Another quantifiable measure lies in the number of citations per category shown in Table \ref{tab: class_counts}. Threat Classification, Detect Threats, and Breach Classification are the dominant fields of study in the previous 10+ years. However, several of the articles in the corpus have exceptionally high citation counts, which skew the numbers. \cite {jouini2014classification} (434), \cite {seh2020healthcare} (532), and \cite {wernke2014classification} (408) affect Threat Classification, Detect Threats, and Breach Classification categories.  After removing these three articles from the citation count and looking at the remaining 26 articles (with individual citation counts of less than 100), the final column in Table \ref{tab: class_counts} paints a different picture. Threat Classification is affected the most, going to 0. The low counts for Predict Breaches (2) and Risk (0) may be attributed to the fact that they are relatively newer studies. 

\begin{table}[htbp]
\centering
\caption{Corpus Classification and Citation Metrics}
\begin{tabular}{|l|c|c|c|}
\hline
\textbf{Classification} & \textbf{Citation Count} & \textbf{$<$100 Citation Count} \\
\hline
Risk & 0 & 0 \\
\hline
Threat Classification & 430 & 0 \\
\hline
Report Classification & 83 & 83 \\
\hline
Predict Breaches & 2 & 2\\
\hline
Detect Threats & 537 & 120 \\
\hline
Detect Breaches & 94 & 94 \\
\hline
Breach Classification & 547 & 15 \\
\hline
\end{tabular}

    \label{tab: class_counts}
\end{table}

\subsection{RQ2: Underexplored \& Emerging Areas}
\begin{figure}[!h]
    \centering
    \includegraphics[width=0.95\linewidth]{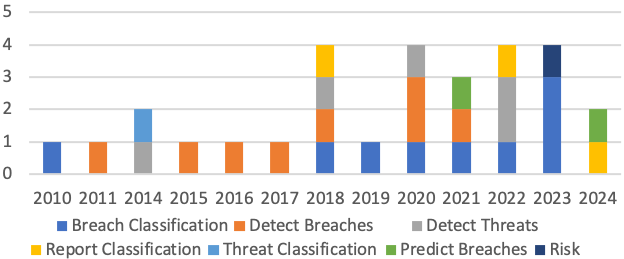}
    \caption{Publications Classified by Derived Categories by Year}
    \label{fig: micro-categories-by-year}
\end{figure}

To address RQ2, the finer-grained taxonomy from Table \ref{tab:taxonomy} was employed. Looking at Fig. \ref{fig: micro-categories-by-year}, Breach Classification remained an area of interest, with the majority of studies appearing in 2023. The Predict Breaches category did not appear until 2021 and again in 2024, indicating that it is an emerging and understudied area of research. Risk also made its first (and only) appearance in 2023, indicating an emerging field of study. Table \ref{tab: class_counts} reinforces the idea that Predict Breaches and Risk represent emerging trends in their respective fields.

\subsection{RQ3: Keyword Frequency}
To address RQ3, word frequency analysis was performed on the selected corpus. Word frequency analysis refers to the number of times words are found in each document. TF-IDF provides a measure of potential gaps in words as shown in Table \ref{tab: word_freq}, or phrases as shown in Table \ref{tab:top_phrase_freq}. The analysis of TF-IDF is important because calculated scores can uncover themes by identifying major words or phrases. Potential gaps can also be identified by the absence of particular topics from the analysis. Table \ref{tab: word_freq} reveals the most important keywords, while Table \ref{tab:top_phrase_freq} reveals the important keyword phrases

\begin{figure}
    \centering
   \includegraphics[width=0.8\linewidth]{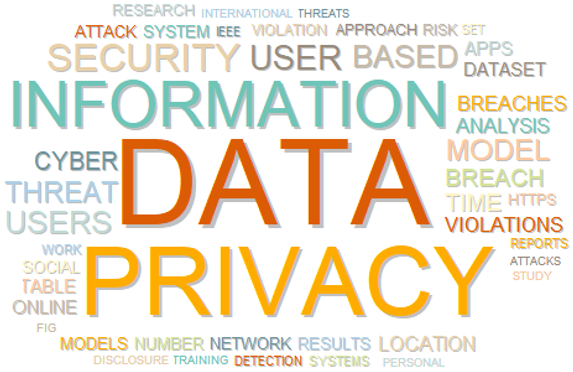}

    \caption{Corpus Word Frequency Cloud}
    \label{fig: word_cloud}
\end{figure}

The word Privacy occurred 1689 times in 27 of 29 documents in the corpus. The term Breach occurred 339 times across 21 of 29 documents. The word Violations occurred 375 times in 13 of 29 documents. Fig. \ref{fig: word_cloud} represents the word frequency cloud for the study corpus. The results are not surprising considering the field and search terms that were used to identify the corpus. It is interesting to note that the terms Model and Threat appear more frequently, with Threat being more meaningful for future investigation.

\begin{table}[ht]
     \caption{Corpus Word Frequency Top Hits}
\begin{adjustbox}{width=\columnwidth,center}
    
    \begin{tabular}{|c|c|c|c|c|c|c|c|}

      \hline
      \textbf{Word} & \textbf{Frequency} & \textbf{\% Shown} & \textbf{\% Processed} & \textbf{\% Total} & \textbf{No. Cases} & \textbf{\% Cases} & \textbf{TF-IDF} \\
        
    \hline 
        Data & 2097 & 3.34\% & 1.90\% & 1.06\% & 29 & 100\% & 0.0 \\
      
        \hline
        \rowcolor{lightgray}
        Privacy & 1689 & 2.69\% & 1.53\% & 0.86\% & 27 & 93.10\% & 52.4 \\
       
        \hline
        Information & 1108 & 1.76\% & 1.00\% & 0.56\% & 29 & 100\% & 0.0 \\
        
        \hline
        Security & 659 & 1.05\% & 0.60\% & 0.33\% & 28 & 96.55\% & 10.0 \\
        
        \hline
        User & 581 & 0.92\% & 0.53\% & 0.29\% & 25 & 86.21\% & 37.5 \\
        
        \hline
        Based & 534 & 0.85\% & 0.48\% & 0.27\% & 29 & 100\% & 0.0 \\
        
        \hline
        Users & 533 & 0.85\% & 0.48\% & 0.27\% & 26 & 89.66\% & 25.3 \\
        
        \hline
        Model & 510 & 0.81\% & 0.46\% & 0.26\% & 27 & 93.10\% & 15.8 \\
        
        \hline
        Threat & 494 & 0.79\% & 0.45\% & 0.25\% & 16 & 55.17\% & 127.6 \\
        
        \hline
        Cyber & 435 & 0.69\% & 0.39\% & 0.22\% & 17 & 58.62\% & 100.9 \\
        
        \hline
        \rowcolor{lightgray}
        Breach & 404 & 0.64\% & 0.37\% & 0.20\% & 18 & 62.07\% & 83.7 \\
        
        \hline
        Time & 387 & 0.62\% & 0.35\% & 0.20\% & 27 & 93.10\% & 12.0 \\
        
        \hline
        \rowcolor{lightgray}
        Violations & 375 & 0.60\% & 0.34\% & 0.19\% & 13 & 44.83\% & 130.7 \\
        
        \hline
        Breaches & 339 & 0.54\% & 0.31\% & 0.17\% & 21 & 72.41\% & 47.5 \\

    \hline
    \end{tabular} \\

\end{adjustbox}
    
    \label{tab: word_freq}
    
\end{table}

The frequency of phrases extracted from the corpus illuminates themes and potential areas for future attention, as shown in Table \ref{tab:top_phrase_freq}. High TF-IDF scores, e.g., Privacy Violations (106.9), indicate phrases that occur frequently in specific documents but are relatively rare across the broader corpus. Such findings highlight their significance within individual studies and their underrepresentation elsewhere, making them potential research gaps worthy of further exploration. A threshold of 60 for TF-IDF was used to identify the phrases Location Based (101.5), Healthcare Data (96.4), Data Provider (89.2), Prediction Models (67.4), Healthcare Data Breaches (64.3), and Android Apps (62.7), and Privacy Policy (67.1) stand out as possible gaps within the corpus.

\begin{table}[ht]
    \caption{Corpus Phrase Frequency Top Hits} 
    
\begin{adjustbox}{width=\columnwidth,center}

    \begin{tabular}{|c|c|c|c|c|}
  
      \hline
             
        \textbf{Phrase} & \textbf{Frequency} & \textbf{No. Cases} & \textbf{\% Cases} & \textbf{TF-IDF} \\
        
    \hline 
        Privacy Violations & 279 & 12 & 41.38\% & \cellcolor{lightgray}106.9 \\
      
        \hline
        Data Breaches & 153 & 14 & 48.28\% & 48.4 \\

        \hline
        Data Breach & 148 & 13 & 44.83\% & 51.6 \\

        \hline
        Location Based & 119 & 4 & 13.79\% & \cellcolor{lightgray} 101.5 \\
        
        \hline
        Privacy Violation & 115 & 9 & 31.03\% & 58.4 \\

        \hline
        International Conference & 112 & 22 & 75.86\% & 13.4 \\
        
        \hline
        Machine Learning & 91 & 17 & 58.62\% & 21.1 \\

        \hline
        Big Data & 84 & 12 & 41.38\% & 32.2 \\

        \hline
        Healthcare Data & 83 & 2 & 6.90\% & \cellcolor{lightgray} 96.4 \\

        \hline
        DOI Org & 82 & 13 & 44.83\% & 28.6 \\

        \hline
        Privacy Policy & 78 & 4 & 13.79\% & \cellcolor{lightgray} 67.1 \\

        \hline
        Location Privacy & 73 & 3 & 10.34\% & \cellcolor{lightgray} 71.9 \\

        \hline
        Personal Information & 72 & 16 & 55.17\% & 18.6 \\

        \hline
        Data Provider & 61 & 1 & 3.45\% & \cellcolor{lightgray} 89.2 \\

        \hline
        Cyber Breach & 60 & 3 & 10.34\% & 59.1 \\

        \hline
        Physical World & 60 & 2 & 6.90\% & 69.7 \\

        \hline
        Prediction Models & 58 & 2 & 6.90\% & \cellcolor{lightgray} 67.4 \\

        \hline
        Cyber Security & 57 & 10 & 34.48\% & 26.4 \\
        
        \hline
        Data Collection & 57 & 11 & 37.93\% & 24.0 \\
       
        \hline
        Social Network & 55 & 6 & 20.69\% & 37.6 \\

        \hline
        Android Apps & 54 & 2 & 6.90\% & \cellcolor{lightgray} 62.7 \\

        \hline
        Threat Intelligence & 54 & 4 & 13.79\% & 46.5 \\

        \hline
        Social Networks & 53 & 8 & 27.59\% & 29.6 \\

        \hline
        Online Social & 50 & 7 & 24.14\% & 30.9 \\

        \hline
        Information Security & 49 & 9 & 31.03\% & 24.9 \\

        \hline
        Location Based Services & 48 & 2 & 6.90\% & 55.7 \\

        \hline
        Data Providers & 46 & 1 & 3.45\% & \cellcolor{lightgray} 67.3 \\

        \hline
        Decision Tree & 46 & 6 & 20.69\% & 31.5 \\

        \hline
        Threat Data & 46 & 3 & 10.34\% & 45.3 \\

        \hline
        Breach Reports & 45 & 6 & 20.69\% & 30.8 \\

        \hline
        D HTTPS & 45 & 3 & 10.34\% & 44.3 \\

        \hline
        Healthcare Data Breaches & 44 & 1 & 3.45\% & \cellcolor{lightgray} 64.3 \\

        \hline
        Computer Science & 43 & 14 & 48.28\% & 13.6 \\

        \hline
        Privacy Concerns & 43 & 12 & 41.38\% & 16.5 \\

        \hline
        Information Systems & 42 & 15 & 51.72\% & 12.0 \\

        \hline
        Privacy Preserving & 41 & 6 & 20.69\% & 28.1 \\

        \hline
        Cyber Insurance & 40 & 4 & 13.79\% & 34.4 \\

        \hline
        Social Media & 40 & 14 & 48.28\% & 12.7 \\

    \hline
    \end{tabular} \\

\end{adjustbox}

    \label{tab:top_phrase_freq}
    
\end{table}

\section {Future Work}\label{future}
Opportunities exist to learn more about privacy or data breach classification and how it applies to the overall protection of privacy, and to investigate new and underrepresented areas in the domain. Beyond the emerging trends and identified gaps, this study could be expanded to garner deeper insights into privacy breach classification. The novel privacy breach classification taxonomy developed by this research provides additional avenues for discovery. For example, each category could be expanded into additional subcategories such as study type, population, and controls. Conducting additional systematic reviews of the literature by category would afford a more focused delineation of word and phrase frequencies. This could provide a more focused view of emerging trends and potential gaps in specific categories of interest.

\section {Conclusion}\label{conclusion}
With the adoption of privacy laws, identifying emerging trends and gaps related to the classification of privacy breaches (violations) is a meaningful exercise to guide discussions, pedagogy, and future research efforts. The application of a topical systematic review of the literature has created a knowledge base that can help researchers identify areas ripe for future research. The research presented a systematic review of the literature to build a corpus of 29 research articles that resulted in a knowledge base related to privacy breach classifications. A novel taxonomy for privacy breach classification was also introduced to quantify related research trends. Research in this domain is on an upward trend, with a boost occurring in 2018. Analysis reveals that classifying and detecting data breaches has dominated research efforts over the last ten years. Detecting Breaches has remained consistent throughout this time frame, while Breach Classification has received more recent attention. Predicting Breaches and Risk analysis were revealed as emerging trends. In addition, word and phrase frequency analysis has identified the following gaps: Android Apps, Data Provider Violations, Location Privacy, Prediction Models, HealthCare Data Breaches, and Privacy Concerns. Emerging trends and identified gaps provide many avenues and opportunities for future work.

\balance
\printbibliography

\end{document}